\documentclass[twocolumn,aps,prd,showpacs,nofootinbib,superscriptaddress,preprintnumbers,floatfix,10pt]{revtex4-1}

\usepackage{amsmath}
\usepackage{amsfonts}
\usepackage{float} 
\usepackage{amssymb}
\usepackage{epsfig}
\usepackage{graphicx}
\usepackage{euscript}
\usepackage{color}
\usepackage{slashed}
\usepackage{epstopdf}
\usepackage[utf8]{inputenc}
\usepackage[normalem]{ulem}

\usepackage{hyperref}

\usepackage{stackrel,amssymb,amsmath}

\newcommand{\leftrarrows}{\mathrel{\raise.75ex\hbox{\oalign{%
  $\scriptstyle\leftarrow$\cr
  \vrule width0pt height.5ex$\hfil\scriptstyle\relbar$\cr}}}}
\newcommand{\lrightarrows}{\mathrel{\raise.75ex\hbox{\oalign{%
  $\scriptstyle\relbar$\hfil\cr
  $\scriptstyle\vrule width0pt height.5ex\smash\rightarrow$\cr}}}}
\newcommand{\Rrelbar}{\mathrel{\raise.75ex\hbox{\oalign{%
  $\scriptstyle\relbar$\cr
  \vrule width0pt height.5ex$\scriptstyle\relbar$}}}}

\makeatletter
\def\leftrightarrowsfill@{\arrowfill@\leftrarrows\Rrelbar\lrightarrows}
\newcommand{\xleftrightarrows}[2][]{\ext@arrow 3399\leftrightarrowsfill@{#1}{#2}}
\makeatother

\newcommand{\be}{\begin{equation}}
\newcommand{\ee}{\end{equation}}
\newcommand{\bea}{\begin{eqnarray}}
\newcommand{\eea}{\end{eqnarray}}

\newcommand{\lgCyclicSixI}[1]{{g}^{3,i}_{6,1}}

\begin{document}

\title {Universal critical behavior in tensor models for four-dimensional quantum gravity
}

\author{Astrid Eichhorn}
\email[]{eichhorn@cp3.sdu.dk} 
\affiliation{CP3-Origins, University of Southern Denmark, Campusvej 55, DK-5230 Odense M, Denmark}
\affiliation{Institute for Theoretical Physics, University of Heidelberg, Philosophenweg 16, 69120 Heidelberg, Germany}

\author{Johannes Lumma}
\email[]{j.lumma@thphys.uni-heidelberg.de} 
\affiliation{Institute for Theoretical Physics, University of Heidelberg, Philosophenweg 16, 69120 Heidelberg, Germany}

\author{Antonio D.~Pereira}
\email[]{adpjunior@id.uff.br}
\affiliation{Instituto de F\'isica, Universidade Federal Fluminense, Campus da Praia Vermelha, Av. Litor\^anea s/n, 24210-346, Niter\'oi, RJ, Brazil}
\affiliation{Institute for Theoretical Physics, University of Heidelberg, Philosophenweg 16, 69120 Heidelberg, Germany}

\author{Arslan Sikandar}
\email[]{a.sikandar@thphys.uni-heidelberg.de}
\affiliation{Physics Department, Quaid-i-Azam University, Islamabad 45320, Pakistan}
\affiliation{Institute for Theoretical Physics, University of Heidelberg, Philosophenweg 16, 69120 Heidelberg, Germany}

\begin{abstract}
Four-dimensional random geometries can be generated by statistical models with  rank-4 tensors as random variables. These are dual to discrete building blocks of random geometries.
We discover a potential candidate for a continuum limit in such a model by employing background-independent coarse-graining techniques where the tensor size serves as a pre-geometric notion of scale. A fixed point candidate which features two relevant directions is found. 
The possible relevance of this result 
in view of universal results for quantum gravity and a potential connection to the asymptotic-safety program is discussed. 
\end{abstract}


\maketitle

\section{
 The case for universal, background-independent quantum gravity}

To describe physically relevant spacetimes, such as  black holes or Friedmann-Lemaitre-Robertson-Walker spacetimes, it is necessary to go beyond General Relativity, where these spacetimes feature singularities. Such singularities are expected to be resolved once quantum fluctuations of spacetime are properly accounted for. 
Yet, this is a particularly challenging task if it is to be compatible with the background-independence at the heart of our modern understanding of gravity. Background-independence implies that no configuration of spacetime should be singled out a priori from all configurations that enter the path integral. 
This is incompatible with perturbative techniques around a fixed spacetime, which single out a special background to perturb around, and which do not provide a predictive quantum field theory of gravity. Instead, one is led to introduce an infinite number of independent local counterterms to cancel divergences at each loop order \cite{tHooft:1974toh,VanNieuwenhuizen:1977ca,Goroff:1985sz}. 
This motivates that background independence should be taken seriously in the search for an ultraviolet complete definition of the gravitational path integral. Thus, we aim at an implementation of the path integral without auxiliary geometric background structures \footnote{An alternative route makes use of an auxiliary background structure at the technical level while ensuring the independence of physical results from this background structure, see, e.g., \cite{Becker:2014qya}.}.
A promising route to construct a background independent path integral consists of making a transition to discrete building blocks, as in dynamical triangulations \cite{Loll:2019rdj}, Regge calculus \cite{Williams:1900zz}, matrix/tensor models \cite{DiFrancesco:1993cyw,Rivasseau:2011hm,Gurau:2016cjo}, spin foams \cite{Perez:2012wv} and causal sets \cite{Surya:2019ndm}. This allows to construct a discrete approximation of all random geometries (and potentially additional configurations with no interpretation as a spacetime geometry) that enter the path integral for quantum gravity. These discrete building blocks are typically not viewed as physical, ``fundamental" building blocks of spacetime. Rather, they are  auxiliary, unphysical entities, allowing to define a regularized path integral in analogy to  lattice gauge theories for non-gravitational quantum field theories. One might object that
  quantum spacetime might be fundamentally discrete, which might suggest that the use of discrete building blocks is appropriate at a physical level.
 Yet, this would require us to guess exactly the right, physical form of the discretization and might feature a predictivity problem, see, e.g., the discussion in \cite{Eichhorn:2019xav}. Further, naive discretizations are likely to break spacetime symmetries see, e.g., \cite{Bahr:2009ku,Bahr:2009qc} for discussions of diffeomorphism symmetry in discrete settings. 
Therefore, a \emph{universal continuum limit} is actually a key demand in order to ensure that one is exploring robust predictions of the quantum-gravity model, instead of specialized artefacts of one particular discretization\footnote{A continuum limit does \emph{not} preclude the  emergence of physical or dynamical discreteness, e.g., in the spectra of certain geometric operators. These different notions of discreteness should not be mixed, and in particular the presence of physical discreteness cannot be used to infer that one should not be taking a continuum limit at the level of (unphysical) configurations in the path integral.}.
Thus, in this framework it is insufficient to search for the continuum-approximation (in the sense of $\ell_{\rm disc}<<\ell$, where $\ell_{\rm disc}$ is the discreteness scale and $\ell$ is the physical distance scale of interest). The continuum approximation will carry non-universal imprints of the details of the theory at $\ell_{\rm disc}$, in the form of contributions which are functions of the finite ratio $\ell_{\rm disc}/\ell$ and which typically come with infinitely many free parameters. In contrast, a universal continuum limit, in which $\ell_{\rm disc}/\ell \rightarrow 0$, ensures that the details of the discretization do not matter and the physics of the model depends on only finitely many free parameters. This is exactly the spirit in which lattice field theories are set up in non-gravitational field theories. We argue that robust physical predictions should be extracted from the gravitational path integral in a similar manner.

\section{Tensor models as a framework for background-independent quantum gravity}
\begin{figure}
\includegraphics[width=\linewidth,clip=true, trim=0cm 5cm 0cm 0cm]{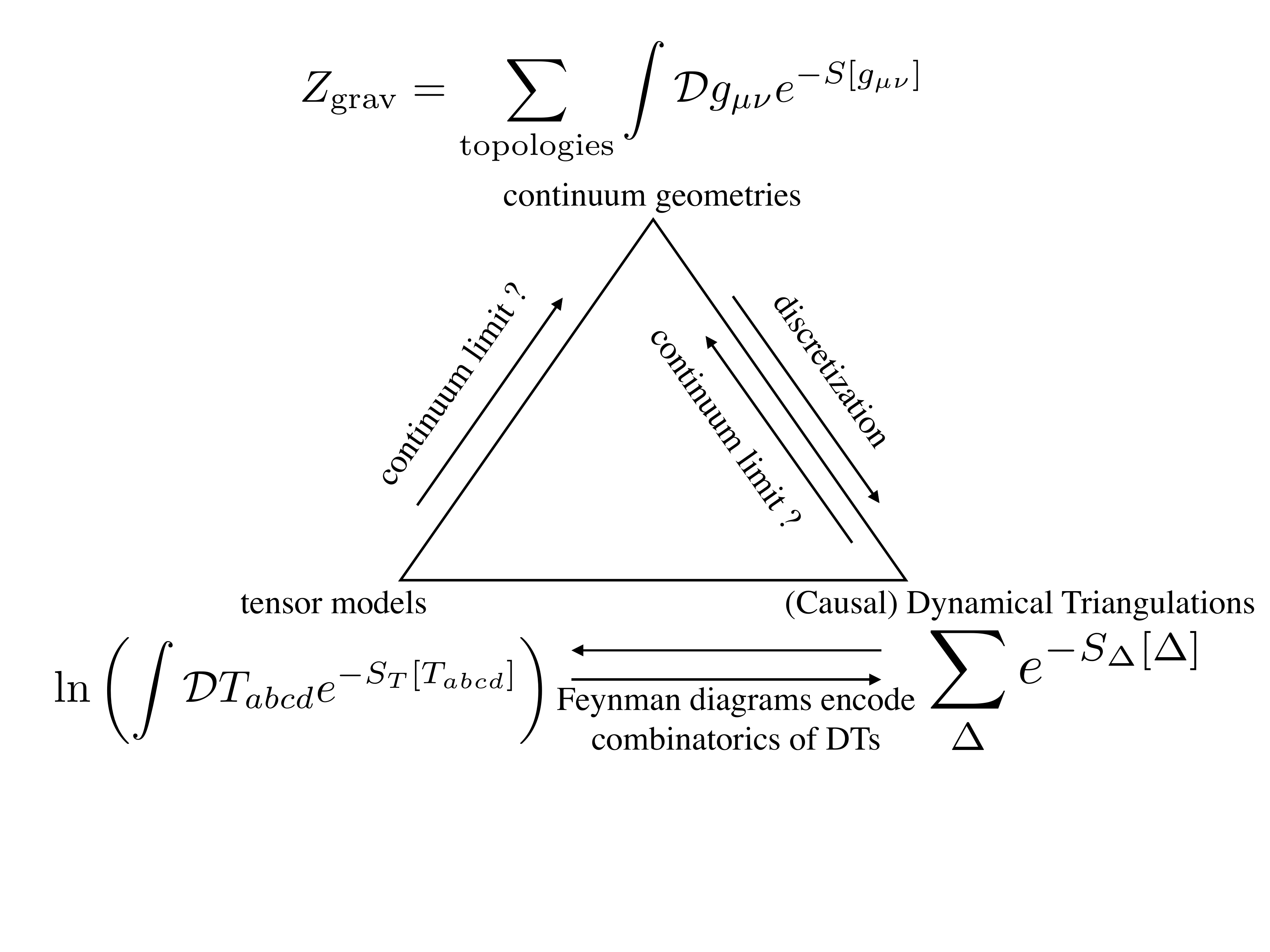}
\caption{\label{fig:illustration_contlim} We show the relation between the generating functional for quantum gravity in the continuum, dynamical triangulations and tensor models.}
\end{figure}

One potential route to evaluate the effect of quantum gravitational fluctuations in a background-independent setting is provided by the tensor-model approach \cite{Ambjorn:1990ge,Godfrey:1990dt,Sasakura:1990fs, Gurau:2009tw,Gurau:2011aq,Gurau:2011xq, Bonzom:2011zz,Bonzom:2012hw,Carrozza:2015adg}.
A tensor model is a zero-dimensional quantum field theory of rank $d$ tensors whose indices range from 0 to $N'$.
The relation between tensor models and the Euclidean gravitational path integral is  provided by the mapping  illustrated in Fig.~\ref{fig:illustration_contlim}.
The first step maps the path integral over all spacetimes to the sum over all triangulations $\Delta$ by discretizing all configurations in terms of discrete, equilateral building blocks, such as, e.g., triangles in two dimensions, tetrahedra in three dimensions and so on. Geometric information is encoded in the way that these building blocks are glued to each other, following the spirit of Regge gravity \cite{Regge:1961px}. For instance, in two dimensions, curvature is encoded in the deficit angle around a vertex, which is only zero if there are exactly six triangles surrounding the vertex. 
The inverse map, from the dynamical triangulations to the path integral for quantum gravity, is via the continuum limit.
Searching for the continuum can be tackled, e.g., by Monte Carlo simulations \cite{Agishtein:1991cv,Catterall:1994pg,Ambjorn:2000dv,Ambjorn:2001cv,Laiho:2011ya,Ambjorn:2011cg,Ambjorn:2012ij,Ambjorn:2017tnl} at the level of the triangulation.\\
Alternatively, one can go one step further to arrive at a purely combinatorial model. In such a tensor model, the building blocks of space are mapped to tensors. The way in which discrete building blocks are glued together to form a discretized configuration of spacetime is encoded in the tensorial interaction structure. More specifically, the Feynman diagram expansion of the tensor model generates all possible triangulations  of pseudo-manifolds \cite{Caravelli:2010nh,Gurau:2011xp}. Such a correspondence between combinatorial models and the path integral for quantum gravity  constitutes a success-story in two dimensions with matrix models (see \cite{DiFrancesco:1993cyw} for a review). Tensor models, first proposed in \cite{Ambjorn:1990ge,Godfrey:1990dt,Sasakura:1990fs}, generalize matrix models to higher dimensions. \\
\begin{figure}[!t]
\centering
\includegraphics[width=0.4\linewidth,clip=true, trim=1.8cm 9.5cm 23.5cm 6cm]{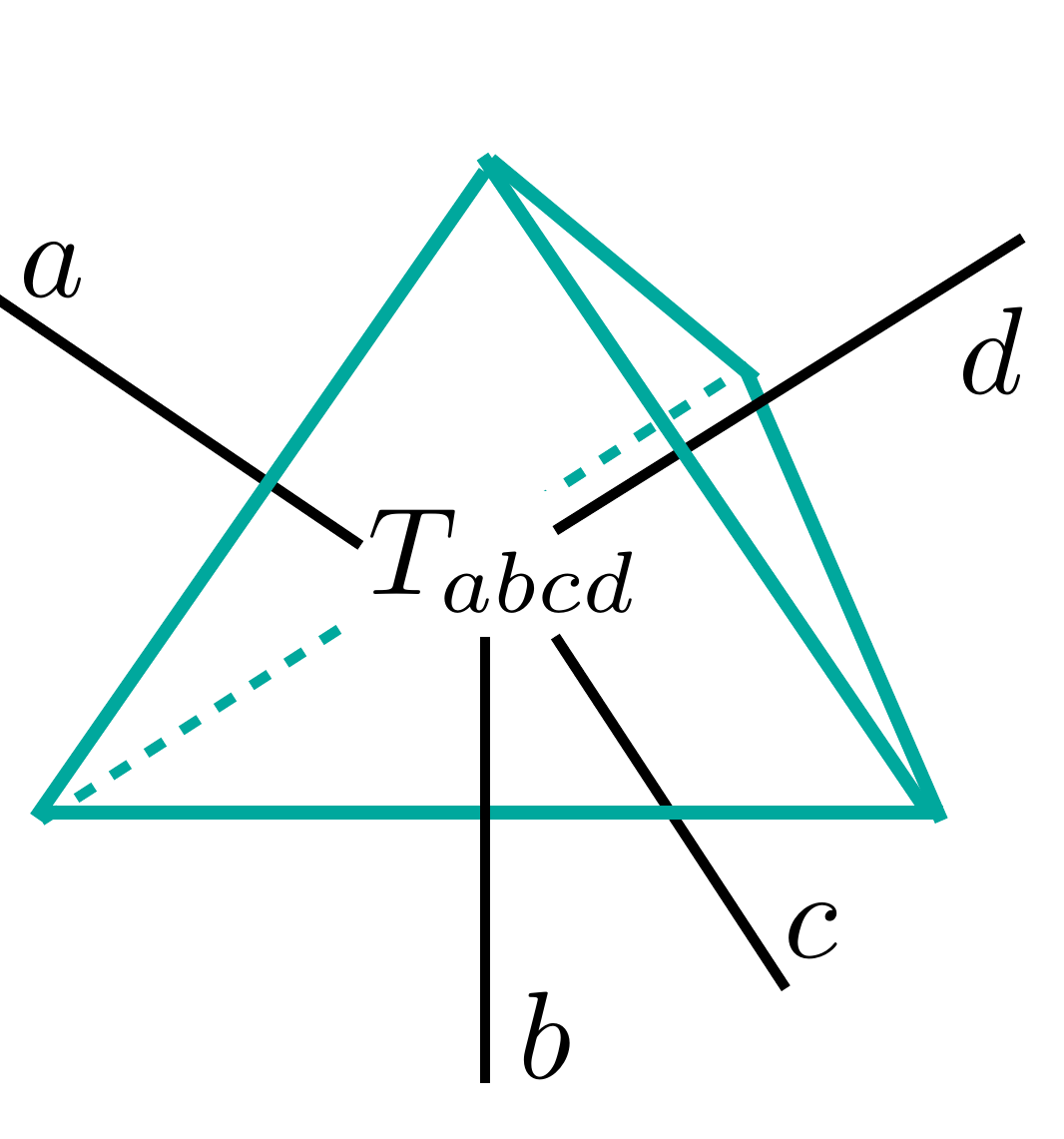}
\caption{\label{fig:tensorill}We show the correspondence between a tensor and a building block of space.}
\end{figure}

\begin{figure}[!t]
\includegraphics[width=0.8\linewidth,clip=true, trim=2cm 6cm 13cm 7cm]{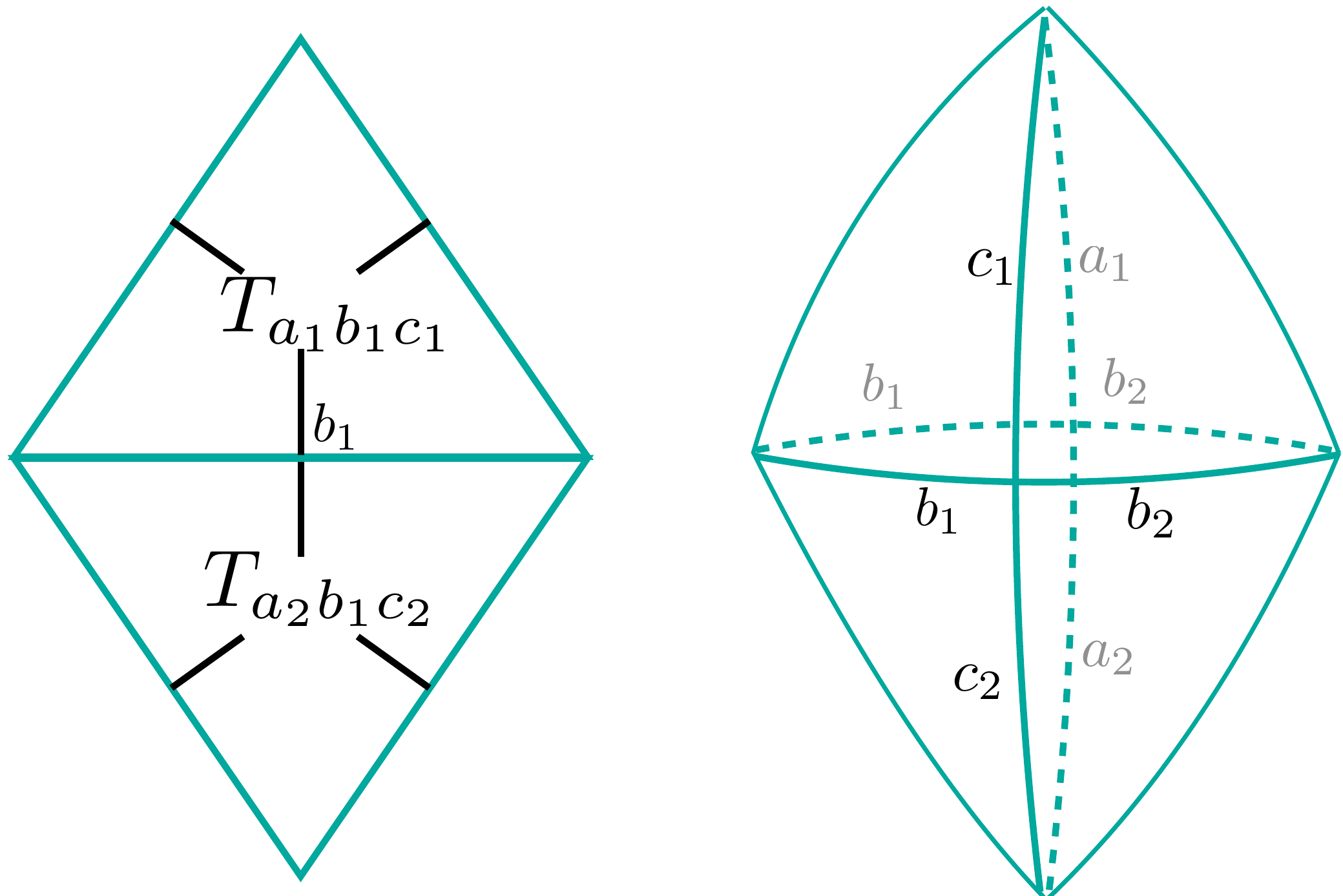}
\caption{\label{fig:melon}The interaction $T_{a_1 b_1 c_1}T_{a_2 b_1 c_2}T_{a_2 b_2 c_2}T_{a_1 b_2 c_1}$ corresponds to a gluing of four triangles to form a ``chunk" of three-dimensional space. We show the corresponding building block of space viewed ``from the front" and ``from the side". The identification of indices of two tensors corresponds to identification of the corresponding edges of the corresponding two triangles. This interaction is known as a ``melonic" one in the literature due to the possible association of a melon with the right-hand diagram.}
\end{figure}

 In detail, the mapping between a Feynman diagram in a tensor model and a triangulation works as follows: For each Feynman diagram of the tensor model, one can construct the dual: Each of the four indices of a tensor is associated to one side of a tetrahedron, cf.~Fig.~\ref{fig:tensorill}. This forms a building block of three-dimensional space. To construct a building block of four-dimensional space, several tetrahedra have to be glued together. This is encoded in a contraction of indices of tensors.
For instance, $T_{a_1b_1c_1d_1}T_{a_1b_1c_1d_2}T_{a_2b_2c_2d_2}T_{a_2b_2c_2d_1}$ encodes how neighbouring tetrahedra are glued together along three/one triangles in a pairwise fashion. For purposes of illustration, we show the analogous construction in one dimension lower in Fig.~\ref{fig:melon}. In summary, the \emph{interactions} of the rank-4-tensor model are dual to building blocks of four-dimensional space, whereas the tensors themselves are dual to building blocks of three-dimensional space. In the Feynman diagram expansion of the tensor model, these interactions are glued together along propagators, forming triangulations of four-dimensional space. In this way, tensor models encode the discretized configurations of spacetime that enter the path-integral in a combinatorial way.

The action $S_T$ only depends on tensor invariants (under orthogonal/unitary transformations of the tensors), such as, e.g., $T_{abcd}T_{abcd}$, with summation over repeated indices implied, but the tensors are \emph{not} functions of spacetime. In this sense, such tensor models are  pre-geometric models. One might wonder how such a simple setting can be rich enough to reproduce the intricacies of quantum gravity. The answer lies in the observation that each equilateral triangulation of space is fully determined by how its building blocks are glued to each other. This ``gluing" information  
is combinatorial information and random tensors are an efficient way of capturing it. At finite $N'$, the tensor models encode the finitely many degrees of freedom of regularized quantum gravity, i.e., discrete random geometries.
In the $N' \rightarrow \infty$ limit, the infinitely many local degrees of freedom of quantum gravity are expected to emerge. A well-defined $1/N'$ expansion is therefore a key prerequisite for the recovery of gravity from this approach. In \cite{Gurau:2010ba,Gurau:2011xp,Gurau:2009tw,Gurau:2011aq,Gurau:2011xq,Bonzom:2012hw,Carrozza:2015adg}, it was shown that a class of models, the (un)colored models, admits a $1/N'$ expansion.
For such models, the interactions are invariant under independent orthogonal (unitary) transformations of each index of the real (complex) tensors. In this paper, we restrict our attention to this subclass of tensor models.

Studies of the so-called loop equations enabled by the $1/N'$ expansion seem to indicate a continuum limit for tensor models which resembles a branched-polymer phase and it is thereby unphysical \cite{Bonzom:2011zz,Gurau:2013cbh}. It has been shown that one can go beyond the branched-polymer phase to recover the continuum limit of 2d-quantum gravity from tensor models \cite{Bonzom:2015axa}. This is tied to a modified scaling of interactions with $N'$, that allows to recover different phase structures \cite{Bonzom:2015axa,Lionni:2017xvn}.
In order to search for a four-dimensional universal continuum limit, it seems necessary to explore novel critical points in which not only $N^\prime$ is taken to be large, but also a set of couplings is tuned to a critical value and interactions exhibit non-canonical scaling with $N'$.  Following the development of functional renormalization group tools for discrete models in \cite{Eichhorn:2014xaa,Eichhorn:2017xhy},  different continuum limits were charted with these techniques also in the related group-field theories \cite{Benedetti:2014qsa,Benedetti:2015yaa,Geloun:2015qfa,Geloun:2016qyb,Geloun:2016xep,Carrozza:2016tih,BenGeloun:2018ekd,Lahoche:2016xiq,Lahoche:2016xiq,Lahoche:2018ggd,Lahoche:2018oeo,Lahoche:2018vun,Carrozza:2017vkz,Lahoche:2018hou,Lahoche:2019vzy,Lahoche:2019cxt,Eichhorn_2019,Lahoche:2019orv}.
For reviews see \cite{Carrozza:2016vsq,Eichhorn:2018phj}. In the present paper, we provide  first hints for such continuum limits for real rank-4  models by discovering a potential universality class that is new from a tensor-model point of view but appears to be \emph{not incompatible} -- within the respective systematic uncertainties -- with results for the Reuter universality class in quantum gravity.

 \section{Universality and the link between tensor models and asymptotically safe gravity}
 A key property of a viable continuum limit is universality. In the sense of statistical physics, it implies the independence of physical results from unphysical microscopic choices.
In quantum-gravity models based on regularizations in terms of discrete building blocks, this implies the independence of the continuum limit from the choice of building blocks (at least within certain classes, defined, e.g., by the emergent symmetries). This implies that certain classes of tensor models should all encode the same continuum limit. For instance, as in \cite{Bonzom:2015axa}, in \cite{Eichhorn_2019}, a certain type of large $N'$ critical behavior of rank-3-models has been found to agree (within the estimated systematic errors) with that of matrix models for two-dimensional quantum gravity through a form of dimensional reduction in which the universal continuum limit becomes independent of the microscopic dimensionality of the building blocks. Yet, universality implies even more, namely that the same universality class can be obtained from a discrete model as well as from a continuum setting. To elaborate this point, let us use the analogy of Yang-Mills theory, which can be explored on the lattice or in the continuum. The lattice spacing corresponds to a UV cutoff which can be imposed equivalently as a momentum cutoff in the continuum. A Renormalization Group (RG) fixed point implies the existence of a universal continuum limit. Specifically, on the lattice side one tunes the relevant bare couplings (i.e., the relevant couplings evaluated at the lattice scale) to the critical surface of the fixed point, such that the physics becomes that of an RG trajectory that can be extended arbitrarily far into the UV, i.e., towards vanishing lattice spacing. The continuum limit is thus enabled by an RG fixed point which can equally well be uncovered by using continuum RG techniques. For quantum gravity, a  similar relationship is expected to hold: If a universal continuum limit can be taken in (causal) dynamical triangulations, this is enabled by an RG fixed point which should be discoverable using continuum RG techniques. Tensor models allow to search for this fixed point on the discrete side, but in a completely pre-geometric, combinatorial setting, where $N'$ provides the notion of scale. The arguable simplest hypothesis about the form of the continuum RG fixed point that a potential large $N'$ fixed point in tensor models corresponds to is that of the Reuter universality class, underlying the asymptotic-safety program, see \cite{Reuter:2012id,Percacci:2017fkn, Eichhorn:2017egq,Eichhorn:2018yfc,Reuter:2019byg} for recent reviews and introductions.\\
Three comments are in order about this hypothesis:  Firstly, in the example of Yang-Mills theory both, the discrete as well as the continuum description, feature the same gauge symmetry. In our scenario, it is not immediately obvious whether the continuum limit in tensor models automatically recovers diffeomorphism invariance. For instance, lattice gravity might also feature a continuum limit determined by the smaller group of foliation-preserving diffeomorphisms. Similarly, on the continuum side, results are actually obtained in a gauge-fixed setting. Secondly, there is the question of topological fluctuations, which are typically to some extent included in tensor models (although they might be suppressed in the large $N'$ limit), but are not expected to be included in the continuum path integral. This potential mismatch  has to be addressed. We stress that a comparison of universality classes obtained in the two settings implicitly provides information about the importance of topological fluctuations for the continuum limit. More generally, the configuration spaces that the path integral is defined on in the discrete and the continuum setting, respectively, should, if they are not in one-to-one agreement, at least include the same effective degrees of freedom relevant for the continuum limit.
Thirdly, the agreement in the dimensionality -- an important ingredient to ensure universality in field-theoretic settings -- might be more subtle in quantum gravity: The microscopic dimensionality associated with the discrete building blocks need not correspond to the dimensionality of the emergent spacetime. The branched-polymer phase is a good example, as it leads to a Hausdorff dimension of 2 and can in fact be obtained from tensor models of various rank (i.e., various, microscopic dimensionality), see \cite{Gurau:2013cbh}. Additionally, the comparison of dimensionality is subject to additional subtleties in quantum gravity, as
different notions of dimensionality need not agree in quantum gravity (see \cite{Reuter:2011ah} for an example) and a scale-dependent notion of dimensionality is generally expected across many different approaches to quantum gravity, see \cite{Carlip:2019onx} for a summary. 

Therefore, we formulate our expectation that a universal continuum limit in tensor models should also be accessible in a continuum language. In the simplest scenario, the Reuter fixed point, defining a gravitational universality class in the continuum \cite{Reuter:1996cp,Lauscher:2001ya,Reuter:2001ag,Lauscher:2002sq,Litim:2003vp,Benedetti:2009rx,Falls:2013bv,Dona:2013qba,Becker:2014qya,Ohta:2015fcu,Gies:2016con,Denz:2016qks,Christiansen:2017bsy,Eichhorn:2018ydy,Knorr:2019atm}, 
 see \cite{Reuter:2012id,Percacci:2017fkn, Eichhorn:2017egq,Eichhorn:2018yfc,Reuter:2019byg} for recent reviews and introductions, should be reproducible from tensor models. The open questions highlighted above imply that it is not a priori clear whether a simple rank-four tensor model is sufficient to achieve this or whether one needs to restrict the corresponding configuration space (e.g., with an appropriate multi-tensor model, generalizing the two-matrix model underlying the restricted configuration space of CDTs \cite{Benedetti:2008hc}) by a more intricate choice of tensor model. As the key result of our paper we will find hints for universal critical behavior in tensor models that is \emph{not incompatible} with the Reuter universality class (given the systematic uncertainties on both sides).

\section{Functional Renormalization Group techniques for tensor models}
To discover a universal scaling limit, we use RG techniques adapted to the discrete setting \cite{Eichhorn:2013isa,Eichhorn:2014xaa}, based on the idea in \cite{Brezin:1992yc}. $N'$ serves as our notion of RG scale, with $N' \rightarrow \infty$ constituting the limit of infinitely many degrees of freedom, i.e., the UV limit. 
This notion of scale differs from the standard notion of scale in the RG, where momentum/energy scales are used. These correspond to the implementation of the RG as a coarse-graining procedure in which one averages fluctuations over local ``patches" -- just as in the original block-spin idea. In quantum gravity, momentum scales fluctuate since the metric fluctuates. Therefore, a different notion of coarse-graining is more suitable. In fact, in a strictly background-independent setting, there is no unique local notion of RG scale. Accordingly, a more abstract notion of scale, tied to the number of effective degrees of freedom, is used here. 
It can be motivated by a more abstract view of the block-spin procedure: By averaging fluctuations over local ``patches" and summarizing many microscopic degrees of freedom in effective, macroscopic degrees of freedom,  block-spin takes us from many degrees of freedom in the UV to fewer degrees of freedom in the IR. This notion of coarse-graining is implemented in tensor models by viewing the tensor size $N^\prime$ as the RG scale.

Universal critical behavior at large $N^\prime$ implies that couplings scale with $N^\prime$ in a particular way. We generalize the so-called double-scaling limit from matrix models \cite{Gross:1989vs,Gross:1989aw,Douglas:1989ve,Brezin:1990rb}, where the coupling $g$ is tuned concertedly with $N'$, schematically 
\be
(g-g_{\rm crit})^{\frac{1}{\theta}}N^\prime = \rm const,
\ee
as $g\rightarrow g_{\rm crit}$ and $N^\prime \rightarrow \infty$. 
This equation leads to the interpretation of the tensor size $N^\prime$ as the RG scale, as it resembles in structure the scaling equation $g(k) = g_{\ast} +(k/k_0)^{- \theta}$ in the local RG, where $k$ is a momentum scale, $g_{\ast}$ the fixed-point value and $\theta$ the critical exponent. 
To implement the idea of an RG flow in $N'$ in practice, one could follow \cite{Brezin:1992yc} to explicitly integrate out the outermost ``layers" of tensors in a Gaussian approximation to derive a perturbative RG flow. Instead, we implement the RG procedure more generally by writing an explicit cutoff term into the generating functional and deriving an equation that encodes the change of the effective dynamics with $N$ as in \cite{Eichhorn:2013isa}. To this end, we define
\be
Z_N[J] = \int_{N'} \mathcal{D}T\, e^{-S[T] + J_{abcd}T_{abcd} - \frac{1}{2}T_{abcd}R_N(a,b,c,d)T_{abcd}}\,.
\ee
The regulator function $R_N (a,b,c,d)$ suppresses the integration of the tensors entries with $a+b+c+d<N$ implementing the aforementioned ``integration of layers'' of the tensor. It actually implements an infrared cutoff, which we highlight by distinguishing the RG scale $N$ from the UV cutoff $N'$.
By a modified Legendre transform, we define the flowing action $\Gamma_N$, which agrees with the standard effective action $\Gamma$ in the limit $N \rightarrow 0$. The flowing action $\Gamma_N$ then reads
\begin{equation}
	\Gamma_N[T]=\sup\limits_{\boldsymbol{J}}\left(\boldsymbol{J}\cdot\boldsymbol{T}-\ln Z_N[J]\right) -\frac{1}{2}\boldsymbol{T}R_N(a,b,c,d)\boldsymbol{T},
\end{equation}
where the bold symbols $\boldsymbol{J}$ and $\boldsymbol{T}$ are a shorthand notation for rank-4 tensors.	
The scale dependence of $\Gamma_N$ is encoded in a functional Renormalization Group equation \cite{Eichhorn:2013isa}, closely resembling its continuum counterpart in \cite{Wetterich:1992yh,Morris:1993qb,Ellwanger:1993kk} in structure:
\begin{equation}
	N\partial_N \Gamma_N=\frac{1}{2}\mathrm{Tr}\left[\left(\frac{\delta^2 \Gamma_N}{\delta \boldsymbol{T}\delta \boldsymbol{T}}+\boldsymbol{R}_N\right)^{-1}N\partial_N \boldsymbol{R}_N\right],
	\label{floweq}
\end{equation}
where we omitted the explicit index-dependence of the regulator.  The derivative with respect to $N$ should be understood within the large-$N$ regime. For finite $N$, it must be replaced by finite difference equation.
One of the key features of the above equation is that universality holds at the fixed points of the effective action $\Gamma_N$, where a $1/N$ expansion is possible. 

An equivalent approach to RG flows in tensor models, based on the Polchinski equation, was developed in \cite{Krajewski:2015clk,Krajewski:2016svb}.

Solving Eq.~\eqref{floweq} exactly is equivalent to completely solving the underlying path integral of the model. Therefore, in practice, one needs to set up approximations to derive a solution of Eq.~\eqref{floweq}. 
Controlled approximations can be devised following an ordering principle for the various terms that can occur in $\Gamma_N$.
For instance, in many examples in local quantum field theories, a reliable approximation to the full flowing action is obtained by including all local terms up to some value of the canonical dimension of the associated coupling. The reliability of the results is tested by enlarging such truncations and checking for stability of the results. In the case of tensor models, such an ordering principle is missing \textit{a priori} since one lacks dimensional analysis: the RG parameter $N$ is just a dimensionless number, and the pregeometric nature of tensor models implies the complete absence of a notion of scaling under changes of length/momentum scales. 
Yet, a notion of canonical dimension for a given coupling in a tensor model can be derived, closely tied to the requirement of a well-defined, but non-trivial large $N$ limit. The flow equation \eqref{floweq} allows for the computation of beta function of the couplings, i.e., for the running of each coupling with the RG scale $N$. In general,  this system of beta functions is non-autonomous, i.e., it depends explicitly on $N$. Given the existence of the $1/N$-expansion for (un)colored tensor models, we rescale the couplings $\bar{g}_i=N^{d_i}g_i$ in such a way that the system of beta functions is compatible with the $1/N$-expansion. In particular, this implies that for the rescaled couplings, the rhs of the beta functions can at most scale with $N^0$. 
The powers $d_i$ are thus the ``canonical dimensions'' and the couplings $g_i$ are called dimensionless couplings.  The beta function $\beta_i$ associated to the coupling $g_i$ has the general structure $\beta_i = -d_i\, g+F(g)$, where the first term of the right-hand side corresponds to the canonical scaling of the coupling and $F(g)$ is a function of the couplings of the theory arising from quantum fluctuations.

A universal continuum limit is possible at fixed points of the RG flow. They are characterized by the simultaneous vanishing of all beta functions $\beta_i$. Associated to the fixed points are the  universal critical exponents defined by minus the eigenvalues of the stability matrix given by $\mathcal{M}_{ij} = \partial\beta_i/\partial g_j$. The critical exponents associated to a fixed point define a universality class. Different microscopic descriptions which lead to the same continuum limit belong to the same universality class. Thus, determining the critical exponents of fixed points is crucial to establish an explicit comparison between the continuum limit of tensor models and other formulations of the path integral for quantum gravity, e.g., in the continuum asymptotic-safety framework.\\

\section{The model}

We explore a real rank-4-model, with an $O(N')\otimes O(N')\otimes O(N')\otimes O(N')$ symmetry, i.e., each index can be rotated  independently. Thus, there is no symmetry that connects the distinct index positions of a tensor. This property was crucial to first establish a well-behaved large $N'$ limit, although more recently more general models have been considered, see, e.g., \cite{Benedetti:2017qxl,Carrozza:2018ewt}.
Moreover, the model we consider here features a maximally-enhanced scaling for the non-melonic interactions\footnote{Within the FRG setup, the canonical scaling dimensions are bounded from above, but can be chosen below the upper bound for some classes of interactions, as lower bounds only exist for some couplings. Whether a non-truncated setup features lower bounds for all couplings is an open question.
The choice of canonical dimension in which all upper bounds are saturated appears to agree with what has been called ``maximally-enhanced" scaling. Other scaling choices lead to the decoupling of certain classes of interactions. Let us stress that a decoupling of so-called multitrace interactions is not possible in this way, see the discussion in \cite{Eichhorn_2019}.}. Tensor models featuring enhanced scaling have also been addressed in \cite{Bonzom:2012wa,Bonzom:2015axa,Bonzom:2016dwy}.
We approximate the effective dynamics, i.e., the flowing action $\Gamma_N$, by all terms up to sixth order in the tensors which are compatible with the $O(N^\prime)^{\otimes 4}$ invariance, adding a subset of interactions at eighth order, namely the so-called melonic couplings, see below. In total, we take into account 170 (including the kinetic term) distinct combinatorial structures.
Comparing with \cite{Avohou:2019qrl}, where the authors compute the number of all possible interactions of a rank-4 O(N) model at each interaction order (see \cite{Geloun:2013kta} for the complex case) provides a nontrivial benchmark of our calculation. Most interaction structures single out a preferred index position, such that these combinatorial structures occur in several incarnations which are related by a permutation of the index position. In this case, we assign the same coupling to all combinatorially equivalent interactions. For instance, the so-called (quartic) cyclic melonic interactions $T_{a_1b_1c_1d_1}T_{a_1b_2c_2d_2}T_{a_2b_2c_2d_2}T_{a_2b_1c_1d_1}$, $T_{a_1b_1c_1d_1}T_{a_2b_1c_2d_2}T_{a_2b_2c_2d_2}T_{a_1b_2c_1d_1}$, $T_{a_1b_1c_1d_1}T_{a_2b_2c_1d_2}\cdot $ $ \cdot T_{a_2b_2c_2d_2}T_{a_1b_1c_2d_1}$ and $T_{a_1b_1c_1d_1}T_{a_2b_2c_2d_1}T_{a_2b_2c_2d_2}\cdot $ $ \cdot T_{a_1b_1c_1d_2}$ all come with the same coupling in this model, but nevertheless all have to be included in the effective dynamics. In \cite{Eichhorn:2018phj} index permutation was treated anisotropically. In many cases, color-anisotropic fixed points feature an enhanced symmetry which implies the reduction of the effective dynamics to lower-rank tensor models \cite{Eichhorn_2019}, motivating us to explore the model with isotropy under index permuations. For completeness let us add that all interactions have an even number of fields as a consequence of the $O(N')\otimes O(N')\otimes O(N')\otimes O(N')$-symmetry and no additional symmetry (e.g., a $\mathbb{Z}_2$ symmetry) needs to be imposed.
  
For the computation of the beta functions, we employ a Litim-type regulator \cite{Litim:2001up}, i.e.,
\begin{eqnarray}
R_{N}(a,b,c,d) &=& Z_N\left(\frac{N}{a+b+c+d}-1\right)\nonumber\\
&\times&\theta\left(\frac{N}{a+b+c+d}-1\right)\,,
\label{Litimreg}
\end{eqnarray}
where $Z_N$ denotes the wave-function renormalization. This choice of regulator implies that the anomalous dimension $\eta = - N \partial_N \ln Z_N$ occurs on the right-hand-side of the flow equation. We will set such terms to zero, which is self-consistent within our truncation scheme that requires that $\eta$ cannot be too large. 

We highlight that the introduction of a regulator that depends on $N$ entails a breaking of the  previously referred to orthogonal/unitary  symmetry of the model, implying the existence of non-trivial Ward identities that have been explored, e.g., in \cite{Lahoche:2018ggd,Lahoche:2019cxt}. 

Details on the calculation and additional results of our study will be presented in a forthcoming paper.

\section{Interacting fixed point}
\begin{figure}[!t]
\includegraphics[width=\linewidth]{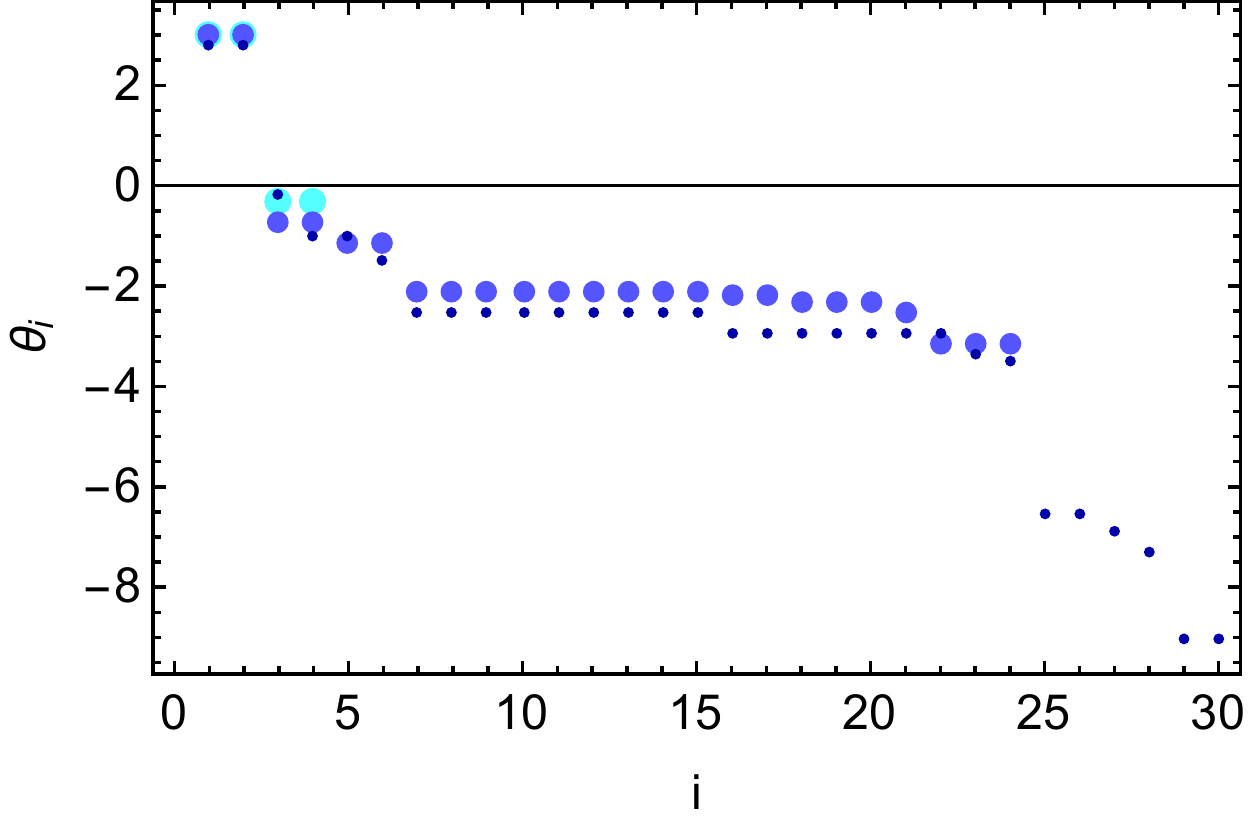}
\caption{\label{fig:theta}We show the real parts of the critical exponents in the $T^4$ (large cyan dots), $T^6$ (medium light blue dots) and $T^8$ (small blue dots) truncation.} 
\end{figure}

We find an interacting fixed which is characterized by two positive critical exponents,
	\be
	\theta_{1,2}=2.79 \pm 1.48\,i.
	\ee
in the $T^8$ truncation.
All remaining critical exponents are negative with $\theta_3=-0.21$ being the negative critical exponent with the smallest absolute value.\\
A key feature of a physical fixed point in distinction to an artefact of the truncation is the behavior under extensions of the truncation. We observe that the leading critical exponents remain reasonably stable, while the anomalous dimension, fourth critical exponent and fixed-point values of the couplings exhibit what one might tentatively interpret as a first hint for the onset of apparent convergence, cf.~Tab.~\ref{tab:fpvalues} and Fig.~\ref{fig:theta}.

Fig.~\ref{fig:theta} exemplifies the self-consistency of our truncation scheme: We work under the assumption that operators which are higher-order in the number of tensors constitute increasingly irrelevant interactions. Our explicit results for the critical exponents follow this expectation. Therefore, we expect that adding higher-order operators to the truncation will not result in additional critical exponents close to $\theta=0$. The total number of relevant directions might nevertheless change under such extensions, since the numerical value of the most relevant critical exponents depends on the (indirect) impact of higher-order operators \footnote{Operators with $n$ tensors only couple directly in beta functions for couplings of interactions with $m\geq n-2$ tensors due to the nonperturbative one-loop structure of the flow equation.}. We provide a tentative estimate of the systematic error induced by our truncation by comparing the change of critical exponents under extensions of the truncation. This results in the tentative conclusion that our estimate for $\theta_3$ is also compatible with a positive third critical exponent.

\begin{widetext}
	\begin{center}
		\begin{table}[!t]
			\begin{tabular}{c||c|c|c|c|c|c|c|c|c|c|c|c|c|c|c}
				truncation & $g_{4,2}^2$ & $g_{4,1}^{2}$&$g^3_{6,3}$& $g^{3}_{6,1}$&$g^{2}_{6,2}$&$g_{8,4}^4$&$g_{8,3}^4$&$g_{8,1}^4$&$g_{8,2}^4$&$g_{8,2,s}^4$&$g_{8,2,m}^4$& $\eta$& $\theta_{1,2}$ & $\theta_3$ & $\theta_4$\\ \hline\hline
				$T^4$ & 11.3 & -1.61 &-&-&-&-&-&-&-&-&-&0.86 &2.996 $\pm$ i 1.227 &-0.288 & -0.288\\\hline
				$T^6$ & 5.36 & -0.98 & 230.1 & -1.42 & -12.43&-&-&-&-&-&- & -0.62 & 2.984 $\pm$ i 1.369 & -0.752 & -0.752\\\hline
				$T^8$ &3.50 &-0.73 &219.6 & -1.68 & -12.29&-300.2&272.8&-2.4&-19.0&-23.6&-6.1 &-0.49 &2.793 $\pm$ i 1.478 & -0.21 & -1.01\\\hline
			\end{tabular}
			\caption{ We find a fixed point characterized by two relevant directions. Within the estimated systematic error it appears possible that the third critical exponent, $\theta_3$, could become positive under extensions of the truncation. {Our notation for the couplings follows that used in \cite{Eichhorn:2017xhy,Eichhorn_2019}: The first lower index denotes the number of tensors in the interaction, e.g., $g_{4}$ denotes all quartic interactions. The second lower index denotes the number of connected components of an interaction. The  upper index indicates the number of ``melonic" parts it includes. Melonic refers to the fact that the interactions are characterized by a summation over three of the four distinct indices of neighbouring tensors, leading to the distinct (1-index)-(3-indices) interaction structure clearly exhibited in Fig.~\ref{fig:FPTruncation}. Additional lower indices denote distinct characteristics of an interaction structure. All other couplings that exist at quartic, hexic and octic order in tensors vanish exactly at the fixed point we explore here.}
			\label{tab:fpvalues}}
		\end{table}
	\end{center}

	\begin{center}
		\begin{figure}[!t]
			\includegraphics[width=1\linewidth,clip=true, trim=0cm 0cm 4cm 0cm]{FixedPointAction.pdf}
			\caption{\label{fig:FPTruncation} We depict all nonvanishing couplings and the corresponding interaction structures at the fixed point. The distinct indices are distinguished by different colors and thick/thin/dashed/dotted lines. All remaining couplings feature a vanishing fixed-point value.}
		\end{figure}
	\end{center}

\end{widetext}

Let us caution that within the non-perturbative approximation (including the full non-polynomial structure of the anomalous dimension), we have not found a real extension of the fixed point found in the largest truncation.  Instead, the fixed point lies at slightly complex fixed-point values in this case. We tentatively conjecture that this is an artefact of the additional zeros of the beta functions that arise in this approximation and can cause fixed-point collisions resulting in complex fixed-point values. We defer further extensions of the truncation, which will provide a check of this conjecture, to future work due to the required technical sophistication to deal with such a large number of interactions.

At the fixed point, only the melonic interactions are non-zero, as shown in Fig.~\ref{fig:FPTruncation}, where all non-zero couplings at the fixed point are depicted together with their corresponding interaction. In  \cite{Bonzom:2011zz,Gurau:2013cbh} is has been discussed that a simple large $N'$ limit dominated by melonic interactions leads to a continuum limit corresponding to the branched-polymer phase, known also from dynamical triangulations. Here, we conjecture that going beyond this limit and exploring interacting fixed point with several relevant directions, such as the present one, might constitute a way to go beyond the branched polymer phase. Yet we caution that probes of the emergent geometry are necessary in order to comprehensively answer this question. In \cite{Lionni:2017xvn}, the complex rank-4 model was studied using a perturbative approach and within a single-trace sector \footnote{We borrow terminology from matrix models, where interactions of the form $M_{ab}M_{bc}M_{cd}M_{da}$ are single-trace, in contrast to $M_{ab}M_{ba}M_{cd}M_{dc}$.  Similarly, we refer to $T_{a_1b_1c_1d_1}T_{a_1b_1c_1d_1} T_{a_2b_2c_2d_2}T_{a_2b_2c_2d_2}$  as a multi-trace interaction, in analogy to the matrix-model case.}. There, no continuum limit was found that corresponds to quantum gravity beyond two dimensions.
At least in the matrix-model case, multi-trace interactions actually encode higher-order curvature terms, see, e.g., \cite{Das:1989fq,Korchemsky:1992tt}. Results in continuum studies of asymptotically safe gravity suggest that higher-order curvature terms are relevant, see, e.g., \cite{Lauscher:2002sq,Codello:2007bd,Machado:2007ea,Benedetti:2009rx,Dietz:2012ic,Falls:2013bv,Falls:2014tra,Falls:2017lst,Falls:2018ylp}.
We therefore tentatively suggest that including multitrace interactions might be important to escape the branched polymer phase.

It is rather intriguing to observe that a gravitational universality class is known which features rather similar -- within the respective systematic errors -- critical exponents to those that we find here, namely the Reuter fixed point underlying asymptotically safe quantum gravity, see, e.g., Tab.~2 in \cite{Eichhorn:2018yfc} for an overview. It also shares the property of a complex pair of relevant critical exponents, found in most truncations, with the present fixed point\footnote{In both cases, it remains to be explored whether the existence of an imaginary part is robust - in particular, it might depend on the parameterization of metric fluctuations  \cite{Gies:2015tca}.}.
In particular, using an exponential parametrization for the continuum metric the authors of \cite{Ohta:2015fcu,Alkofer:2018fxj,deBrito:2018jxt} found a fixed point with two relevant directions to describe the universality class of the asymptotic safety scenario for gravity. 
 Moreover, hints for the existence of a fixed point with two relevant directions were found in a unimodular setting, see \cite{Eichhorn:2015bna}. We reiterate that the systematic uncertainties for $\theta_3$ are too large to robustly conclude that it cannot become positive. The Reuter fixed point and the potential universality class for tensor models we find here appear to be not incompatible, i.e., within the systematic errors we cannot exclude that they are in fact the same universality class. This motivates extended studies aimed at reducing the systematic error in order to be able to make an informed decision about the agreement of the two tentative universality classes.

Drawing direct conclusions about the emergent geometry from the tensor model is challenging. Once agreement with a continuum universality class is established, it is of course much simpler to access the geometric properties through calculations on the continuum side. Reducing the systematic error of studies such as the present one to enable a robust comparison to the Reuter universality class is therefore a promising route to learn about the emergent geometries. Using matter as an additional probe, enabling, e.g., the extraction of a spectral dimension, could be an additional possibility.\newline\\

\noindent\emph{Acknowledgements:}\\
We thank Tim Koslowski, Joseph Ben Geloun, Vincent Lahoche and Luca Lionni for discussions.
ADP thanks FAPERJ for support within the ``Jovem Cientista do Nosso Estado" program.
This work is supported by the DFG under grant no.~Ei-1037/1. This research is also partially supported by the Danish National Research Foundation under grant DNRF:90. A.~E.~is also supported by a visiting fellowship at the Perimeter Institute for Theoretical Physics.

\bibliography{refsTensorLetter}
\end{document}